\documentclass[a4paper,twocolumn,aps,showpacs,showkeys,preprintnumbers,amsmath,amssymb,nofootinbib,superscriptaddress]{revtex4}
\usepackage{epsfig}
\usepackage{amsfonts}
\usepackage{dcolumn}

\newcommand{\newc}{\newcommand}
\newc{\gsim}{\lower.7ex\hbox{$\;\stackrel{\textstyle>}{\sim}\;$}}
\newc{\lsim}{\lower.7ex\hbox{$\;\stackrel{\textstyle<}{\sim}\;$}}
\newc{\gev}{\,{\rm GeV}}
\newc{\mev}{\,{\rm MeV}}
\newc{\ev}{\,{\rm eV}}
\newc{\kev}{\,{\rm keV}}
\newc{\tev}{\,{\rm TeV}}

\newc{\mz}{M_Z}
\newc{\mpl}{M_*}
\newc{\mw}{m_{\rm weak}}
\newc{\nr}[1]{N^c_R{}_{#1}}
\usepackage{amsmath}

\def\beq{\begin{equation}}
\def\eeq{\end{equation}}
\def\bea{\begin{eqnarray}}
\def\eea{\end{eqnarray}}
\def\bi{\begin{itemize}}
\def\ei{\end{itemize}}
\newc{\ie}{{\it i.e.}}          \newc{\etal}{{\it et al.}}
\newc{\eg}{{\it e.g.}}          \newc{\etc}{{\it etc.}}
\newc{\cf}{{\it c.f.}}
\def\vev#1{\left\langle #1 \right\rangle}

\def\abs#1{\left| #1\right|}

\def\inv{^{\raise.15ex\hbox{${\scriptscriptstyle -}$}\kern-.05em 1}}
\def\lbar{{\lower.35ex\hbox{$\mathchar'26$}\mkern-10mu\lambda}} 

\let\<=\langle
\let\>=\rangle

\let\+=\uparrow

\let\ga=\gamma

\let\la=\lambda

\let\si=\sigma

\begin{document}
\title{Non-Perturbative Flat Direction Decay} 
\date{20th July}
\author{Anders Basb\o ll}  
\affiliation{Department of Physics and Astronomy, University of Aarhus,
Ny Munkegade, DK-8000 Aarhus C}

\author{David Maybury}  
\affiliation{Rudolf Peierls Centre for Theoretical Physics, University of Oxford,
1 Keble Rd., Oxford OX1 3NP, UK}

\author{Francesco Riva}  
\affiliation{Rudolf Peierls Centre for Theoretical Physics, University of Oxford,
1 Keble Rd., Oxford OX1 3NP, UK}

\author{Stephen M. West}
\affiliation{Rudolf Peierls Centre for Theoretical Physics, University of Oxford,
1 Keble Rd., Oxford OX1 3NP, UK}

\begin{abstract}
We argue that supersymmetric flat direction vacuum expectation values can decay non-perturbatively via preheating. Considering a toy $U(1)$ gauge theory, we explicitly 
calculate the scalar potential, in the unitary gauge, for excitations around several flat directions. We show that the mass matrix for the excitations has non-diagonal entries which
vary with the phase of the flat direction vacuum expectation value. Furthermore, this mass matrix has zero eigenvalues whose eigenstates change with time. We show that these 
light degrees of freedom are produced copiously in the non-perturbative decay of the flat direction vacuum expectation value.
\end{abstract}   
\pacs{12.60.Jv 98.80.Cq 98.80.-k}
\keywords{Flat directions, Preheating, Supersymmetry}
\preprint{OUTP-0702P}
\maketitle 

\section{Introduction}

The scalar potential of the Minimal Supersymmetric Standard Model (MSSM) possesses a large number of F- and D-flat directions along which the scalar potential nearly vanishes 
\cite{Gherghetta:1995dv,Enqvist:2003gh}. These flat directions can have important cosmological consequences, including the generation of the baryon asymmetry of the Universe 
through the out-of-equilibrium CP violating decay of coherent field oscillations along the flat directions themselves \cite{Affleck:1984fy,Linde:1985gh,Dine:1995uk}.

Recently, much interest has focused on the cosmological fate of flat direction vacuum expectation values (vev)s. In particular, it has been argued \cite{Allahverdi:2006iq} that in 
realistic supersymmetric models, large flat direction vevs can persist long enough to delay thermalization after inflation and therefore lead to low reheat 
temperatures. Furthermore, it has also been asserted \cite{Allahverdi:2006wh} that large flat direction vevs can prevent non-perturbative parametric resonant decay (preheating) 
of the inflaton since the inflaton decay products become sufficiently massive preventing preheating from ever becoming efficient. These arguments hold so long as the flat 
direction vevs do not rapidly decay -- they must persist long enough so that they can delay thermalization and block inflaton preheating. In \cite{Olive:2006uw} it was claimed that 
non-perturbative decay can lead to a rapid depletion of the flat direction condensate and thus precludes the delay of thermalization after inflation. It was also concluded that in 
order for the flat direction to decay non-perturbatively the system requires more than one flat direction \cite{Olive:2006uw,Allahverdi:2006xh}. Finally, in \cite{Allahverdi:2006xh} 
it was pointed out that even in the presence of multiple flat directions, some degree of fine-tuning was necessary to achieve flat direction decay.

An important aspect of this discussion centers on the issue of Nambu-Goldstone (NG) bosons. In general, supersymmetric flat directions are charged under the gauge group of the 
MSSM. Consequently, the flat direction vev will break some or all of the gauge symmetries of the theory and thus we expect the presence of the associated NG bosons. In 
calculating non-perturbative flat direction decays, \cite{Olive:2006uw} considers a gauged $U(1)$ model and constructs the mixing matrix for the excitations around the flat 
direction vev. The results in \cite{Olive:2006uw} show that in the single flat direction case, non-perturbative decay proceeds solely via a massless NG mode as only the NG mode 
mixes with the Higgs and all other massless moduli remain decoupled. Since the NG boson represents an unphysical gauge degree of freedom, it was concluded 
\cite{Olive:2006uw,Allahverdi:2006xh} that no preheating occurs in the single flat direction case. As the appearance of a massless NG boson in the spectrum is a gauge dependent 
artifact, it remains unclear if the conclusions drawn about the system hold in the unitary gauge. In order to determine if flat direction vevs decay non-perturbatively into scalar 
degrees of freedom, the effect of the NG boson mixing in the scalar potential must first be removed. The process of removing the NG modes by switching to the unitary gauge 
changes the form of the mixing matrix among the left over scalar degrees of freedom. 

In this letter we consider toy models to demonstrate that, in the unitary gauge, the mixing matrix of the excitations around a flat direction vev permits preheating. Moreover, we 
find that flat direction decay depends on the \emph{number of dynamical, physical phases} appearing in the flat direction vev. Specifically, a physical phase difference between 
two of the individual field vevs making up the flat direction is needed. 

The outline of the rest of this letter proceeds as follows: first we explicitly construct -- in the unitary gauge -- the mass squared matrix arising from the D-terms of a toy gauged 
$U(1)$ model with three charged chiral superfields. We then present the formalism of preheating with multi-component fields and show that preheating occurs for the light 
moduli associated with the flat direction. We then analyze the specific dynamics of the background field equations for the toy models examined. Finally we evolve the background 
field equations for one of the toy models to obtain quantitative results.

\section{Toy Model with a gauged $U(1)$ Symmetry.}\label{sfd}

As an example, we examine a toy model which demonstrates the important features of supersymmetric flat direction vev decay. We introduce three complex 
superfields\footnote{We use $\Phi$ to denote both superfields and scalar components of superfields.} , 
$\Phi_1$, $\Phi_2$ and $\Phi_3$ charged under a $U(1)$ gauge group with charges $q_1=+1/2$, $q_2=-1$ and $q_3=+1/2$  respectively. The Lagrangian reads 
\begin{equation}\label{Lagrangian}
\mathcal{L}=\sum_{i=1}^3\frac{1}{2}|D_{\mu}\Phi_i|^2-V-\frac{1}{4}F_{\mu\nu}^2,
\end{equation}
where $D_{\mu}=\partial_{\mu}-iq_iA_{\mu}$ denotes the covariant derivative. The potential we consider arises from the supersymmetric D-terms and has the form
\beq
V=\frac{g^2}{2}(q_1\abs{\Phi_1}^2+q_2\abs{\Phi_2}^2+q_3\abs{\Phi_3}^2)^2,
\label{pot3fields}
\eeq
where $g$ is the gauge coupling of the $U(1)$ gauge symmetry. In the above, we have neglected contributions from supersymmetry (SUSY) breaking and from any 
non-renormalisable terms arising from the superpotential. These contributions are highly model-dependent and cloud the analysis we wish to present. A fully realistic model must 
include these additional contributions which can significantly affect the resulting particle production. The effects we investigate here do not depend on their inclusion in the 
quadratic part of the potential and so for clarity we neglect them\footnote{SUSY breaking terms are needed for the computation of the evolution of the flat direction vev.}.

The potential in eq.\ref{pot3fields} admits several flat directions. Choosing one particular direction and including excitations around the vev we can write
 \bea
 \nonumber
 \label{fields21}
 \Phi_1&=&(\varphi+\xi_1)e^{i(\si_1+\frac{\theta_1}{\varphi})}, \nonumber\\
 \Phi_2&=&(\varphi+\xi_2)e^{i(\si_2+\frac{\theta_2}{\varphi})},\\
 \Phi_3&=&(\varphi+\xi_3)e^{i(\si_3+\frac{\theta_3}{\varphi})}\nonumber,
 \eea
where $\sigma_{1,2,3}$ represent time dependent phases of the vevs, $\varphi$ denotes the vev's time dependent amplitude\footnote{Throughout this analysis we assume 
$\dot{\varphi}\ll \varphi\dot{\si}$.} while $\xi_{1,2,3}$ and $\theta_{1,2,3}$ parameterize the six real scalar degrees of freedom corresponding to the excitations around the vevs.
Note that the flat direction vev breaks the $U(1)$ gauge symmetry. Thus, out of the six real scalar degrees of freedom we expect one massive Higgs field and one massless NG 
boson, leaving four massless scalar degrees of freedom.

The kinetic terms for the scalar fields play an important role in this analysis. Their expansion in eq.\ref{Lagrangian} includes the term
\begin{equation}\label{LkinExpansion}
\mathcal{L}\supset -\varphi^2A_0(\dot{\si_1}-2\dot{\si_2}+\dot{\si_3})
\end{equation}
which has the form of a coupling between the gauge field and the background condensate. Terms of this type will feed into the equations of motion for the gauge field which, in 
turn, will have an effect on the equations of motion for the scalar excitations. By making a $U(1)$ gauge transformation on the vev of the form
\beq
\vev{\Phi_i}\rightarrow\vev{\Phi^{\prime}_i}=e^{iq_i\la}\vev{\Phi_i}
\eeq
with
\beq
\la=\frac{2\si_2-\si_1-\si_3}{3},
\eeq
we can gauge this term away and avoid a complicated analysis of the kinetic terms. The resulting form of the vev reads
\bea\label{mod1vev}
\vev{ \Phi_1}&=&\varphi e^{i(\si+\ga)}, \nonumber\\
 \vev{\Phi_2}&=&\varphi e^{i\si},\\
 \vev{\Phi_3}&=&\varphi e^{i(\si-\ga)},\nonumber
\eea
where $\gamma=(\si_1-\si_3)/2$ and $\si=(\si_1+\si_2+\si_3)/3$ represent the two remaining independent physical phases.
Following Kibble \cite{Kibble}, we can write the fields in the unitary gauge as
\begin{eqnarray}\label{fieldsSingleExcitationsNoNG}
\Phi_1 &=& (\varphi +\xi_1) e^{i( \si+\gamma+\frac{\theta}{\varphi\sqrt{2}}+ \frac{\theta^\prime}{\varphi\sqrt{3}})}, \nonumber \\
\Phi_2 &=& (\varphi + \xi_2)e^{i (\si+ \frac{\theta^\prime}{\varphi\sqrt{3}})}, \\
\Phi_3 &=& (\varphi + \xi_3)e^{i (\si-\gamma-\frac{\theta}{\varphi\sqrt{2}}+ \frac{\theta^\prime}{\varphi\sqrt{3}})}, \nonumber
\end{eqnarray} 
where $\xi_{1,2,3}$, $\theta$ and $\theta^\prime$ denote the physical excitations -- the NG boson has been removed\footnote{This can be verified by expanding out the scalar 
kinetic terms which reveals the absence of terms of the form $A_{\mu}\partial^{\mu}(...)$.}. We choose the particular combination of field excitations appearing in 
eq.\ref{fieldsSingleExcitationsNoNG} (the exponent in particular) in order to retain canonically normalised kinetic terms. 

On substituting the fields of eq.\ref{fieldsSingleExcitationsNoNG} into the Lagrangian given in eq.\ref{Lagrangian} and defining the vector
$\Xi\equiv(\xi_1,\xi_2,\xi_3,\theta,\theta^\prime)^T$, we find the quadratic terms
\begin{equation}\label{LagrangianExpanded}
\mathcal{L}\supset\frac{1}{2}|\partial_{\mu}\Xi|^2-\frac{1}{2}\Xi^T \mathcal{M}^2\Xi+\frac{1}{2}\dot{\Xi}^TU\Xi+...
\end{equation}
where the ellipses denote higher order terms and interactions. The matrix $U$ given in the last term in eq.\ref{LagrangianExpanded} reads
\begin{equation}
U= \left(\begin{array}{ccccc} 0 & 0 &0&\frac{\dot{\si}+\dot{\gamma}}{\sqrt{2}}&\frac{\dot{\si}+\dot{\gamma}}{\sqrt{3}} \\ 0& 0  & 0&0&\frac{\dot{\si}}{\sqrt{3}} \\ 
0&0&0&\frac{-\dot{\si}+\dot{\gamma}}{\sqrt{2}}&\frac{\dot{\si}-\dot{\gamma}}{\sqrt{3}}\\-\frac{\dot{\si}+\dot{\gamma}}{\sqrt{2}} 
&0&\frac{\dot{\si}-\dot{\gamma}}{\sqrt{2}} &0&0\\ -\frac{\dot{\si}+\dot{\gamma}}{\sqrt{3}}&-\frac{\dot{\si}}{\sqrt{3}}&\frac{-\dot{\si}+\dot{\gamma}}{\sqrt{2}}&0&0
\end{array} \right)
\end{equation}
while the mass matrix for the physical excitations appears as
\beq\label{MU(1)}
\mathcal{M}^2=(g\varphi)^2\left ( \begin{array}{ccccc} 1 & -2 & 1&0&0\\ -2& 4  & -2&0&0 \\ 1 & -2  & 1 &0&0\\0&0&0&0&0\\0&0&0&0&0   
\end{array} \right)=B\mathcal{M}_d^2B^T
\eeq
with eigenvalues $M^2_1=6(g\varphi)^2$, $M^2_2=M^2_3=M^2_4=M^2_5=0$ (the entries of the diagonal matrix $\mathcal{M}_d$). $B$ is an orthogonal matrix which 
diagonalizes $\mathcal{M}^2$ and $M_1$ corresponds to the mass of the physical Higgs field associated with the spontaneous breaking of the $U(1)$ symmetry. The four zero 
eigenvalues correspond to the massless excitations around the flat direction vev.

The last term in eq.\ref{LagrangianExpanded} appears as a consequence of the time-dependence of the background -- it represents a mixing between the fields $\xi_{1,2,3}$, 
$\theta$, $\theta^\prime$ and their time-derivatives. The effect of these terms on the system becomes clear if we make field redefinitions that remove the mixed derivative terms. 
The resulting transformation leaves the system in an inertial frame in field space and leads to a time-dependent mass matrix. Defining $\Xi^\prime=A\Xi$ ($A$ is orthogonal), 
we find the condition that $A$ must satisfy, in order for all the mixed derivative terms to cancel, to be
\begin{equation}
\dot{A}^TA=U.
\end{equation}
The Lagrangian for the $\Xi^\prime$ system now reads
\begin{equation}
\mathcal{L}\supset \frac{1}{2}|\partial_{\mu}\Xi^{\prime}|^2-\frac{1}{2}\Xi^{\prime T} \mathcal{M}^{\prime 2}\Xi^{\prime}
\end{equation}
where $\mathcal{M}^{\prime 2}=A\mathcal{M}^2A^T=AB\mathcal{M}_d^2B^TA^T=C\mathcal{M}_d^{2}C^T$
and $C=AB$. The matrix $C$ is an orthogonal time-dependent matrix, with columns corresponding to the eigenvectors of $\mathcal{M}^{\prime 2}$. We now have a system of 
scalar fields with canonically normalized kinetic terms and time dependent eigenvectors.

The central point of this discussion centers precisely on the appearance of the {\it time dependent} eigenvectors for the five scalar fields. This satisfies a necessary but not 
sufficient condition for preheating. In the next sections, we investigate the details of the non-perturbative production of the light scalar fields following the analysis of 
\cite{Nilles:2001fg}.

\section{Non-perturbative production of particles}

Including gravity, the dynamics of the re-scaled conformally coupled scalar fields, $\chi_i=a\Xi^{\prime}_i$, where $a$ denotes the scale factor and $\Xi^{\prime}_i$ the $i$-th 
component of the vector $\Xi^{\prime}$, are governed by the following equations of motion (sum over repeated indices is implied),
\begin{equation}\label{eommany}
\ddot{\chi_i}+\Omega^2_{ij}(t)\chi_j=0
\end{equation}
where dots represent derivatives with respect to conformal time $t$, and
\begin{equation}
\label{Omega}
\Omega^2_{ij}=a^2 \mathcal{M^{\prime}}_{ij}^2+k^2\delta_{ij},
\end{equation}
where $k$ labels the comoving momentum. Using an orthogonal time-dependent matrix $C(t)$, we can diagonalize $\Omega_{ij}$ via
$C^T(t)\Omega^2(t)C(t)=\omega^2(t)$, giving the diagonal entries $\omega^2_j(t)$. Terms of the form $\sim {\varphi}\dot{\sigma}\dot{\chi}$ arising from the kinetic terms do 
not affect the evolution of the non-zero $k$ quantum modes \cite{Casadio:2007ip}. 

Once we have identified the basis in which the Hamiltonian appears diagonal (via the orthogonal matrix $C(t)$), the study of particle creation by the time-varying background 
proceeds as in \cite{Traschen:1990sw,KLS,Nilles:2001fg}, which extends the results of \cite{Zeldovich:1971mw}. Following \cite{Nilles:2001fg}, we assume that $\Omega_{ij}$ 
initially evolves adiabatically by assuming that the initial angular motion of the flat-direction varies slowly. This assumption allows us to define adiabatically evolving mode 
functions with positive and negative frequency. We rewrite the quantum fields as mode expansions in terms of the mode functions and their associated creation/annihilation 
operators which allows us to define the initial vacuum. During the evolution, the entries of $\Omega_{ij}$ do not necessarily change adiabatically and consequently we must find 
new mode functions that satisfy eq.\ref{eommany}. A new set of creation/annihilation operators required to define the new vacuum can be related to the initial set using a 
Bogolyubov transformation with Bogolyubov coefficients $\alpha$ and $\beta$ (which denote matrices in the multi-field case). 

Initially $\alpha=\mathbb{I}$ and $\beta=0$ while the coupled differential equations (matrix multiplication implied)
\bea\label{alphadot}
\dot{\alpha} &=& -i \omega\alpha + \frac{\dot{\omega}}{2\omega} \beta - I \alpha - J\beta \nonumber \\
\dot{\beta} &=& \frac{\dot{\omega}}{2\omega} \alpha + i\omega\beta - J \alpha - I\beta
\label{alandbe}
\eea
govern the system's time evolution with the matrices I and J given by
\begin{equation}
I=\frac{1}{2}\left(\sqrt{\omega}\,C^T\dot{C}\frac{1}{\sqrt{\omega}}+\frac{1}{\sqrt{\omega}}\,C^T\dot{C}\sqrt{\omega}\right),
\end{equation}
\begin{equation}\label{jmatrix}
J=\frac{1}{2}\left(\sqrt{\omega}\,C^T\dot{C}\frac{1}{\sqrt{\omega}}-\frac{1}{\sqrt{\omega}}\,C^T\dot{C}\sqrt{\omega}\right).
\end{equation}
Similarly to the single-field case it can be shown \cite{Nilles:2001fg} that at any generic time the occupation number of the $i$th bosonic eigenstate reads
\begin{equation}\label{n}
n_i(t)=(\beta^*\beta^T)_{ii}.
\end{equation}
As pointed out in \cite{Nilles:2001fg,Olive:2006uw}, there exists two sources of non-adiabaticity in the multi-field scenario. The first source arises from the individual frequency 
time dependence and appears as the only source of non-adiabaticity in the single field case. The second source appears from the time dependence of the frequency matrix 
$\Omega_{ij}$ giving rise to terms in eq.\ref{alandbe} proportional to $I$ and $J$. This second source provides the most important contribution in our analysis and gives rise to 
non-perturbative particle production. 

Since initially $\alpha=\mathbb{I}$ and $\beta=0$, eq.\ref{alphadot} shows that a non-vanishing matrix $J$ is a necessary condition to obtain $\dot{\beta}\neq 0$ and hence 
$n_i(t)\neq0$. In general, we have
\begin{equation}
C^T\dot{C}=B^TA^T\dot{A}B=-B^TUB
\end{equation}
where $A$, $B$ and $U$ were defined in the previous section. For the toy $U(1)$ example outlined above, $J$ is a $5\times5$ matrix in the $\chi_i$ basis with non-vanishing 
components
\begin{equation}
J_{1, 2}=J_{2, 1}=\frac{k-\sqrt{k^2+M_1^2}}{2\sqrt{3k}(k^2+M_1^2)^{1/4}}\dot{\gamma},
\end{equation}
where $M_1$ denotes the mass of the heavy Higgs field. These entries in the matrix $J$ link the eigenstate of the Higgs ($i=1$) with one of the light eigenstates ($i=2$). We see that 
in the toy $U(1)$ model, preheating can occur provided that $\dot{\ga}\ne 0$. We address this point in a later section.

\section{Multiple VEV Amplitudes}

We can extend the analysis of the previous sections by allowing the magnitudes of the individual field vevs to differ from one another. As above, we consider the case with three 
complex superfields charged under a $U(1)$ gauge group with charges $q_1=+1/2$, $q_2=-1$ and $q_3=+1/2$ respectively, and with the scalar potential given in 
eq.\ref{pot3fields}. We can write the flat direction with the following vev
\bea
\vev{ \Phi_1}&=&\varphi_1 e^{i\si_1}, \nonumber\\
 \vev{\Phi_2}&=&\frac{1}{\sqrt{2}}(\varphi_1^2+\varphi_2^{2})^{1/2}e^{i\si_2},\\
 \vev{\Phi_3}&=&\varphi_2 e^{i\si_3}\nonumber.
\eea
By substituting the above into the potential given in eq.\ref{pot3fields}, it can readily be shown that the configuration satisfies D-flatness. Expanding around this vev we have
 \bea
 \nonumber
 \label{fields2fdexcitations}
 \Phi_1&=&(\varphi_1+\xi_1)e^{i\left(\si_1+\frac{\theta_1}{\varphi_1}\right)},\nonumber \\
 \Phi_2&=&\left(\frac{1}{\sqrt{2}}(\varphi_1^2+\varphi_2^{2})^{1/2}+\xi_2\right)e^{i\left(\si_2+\frac{\sqrt{2}\theta_2}{\left(\varphi_1^2+\varphi_2^{ 
2}\right)^{1/2}}\right)},\nonumber\\
 \Phi_3&=&(\varphi_2+\xi_3)e^{i\left(\si_3+\frac{\theta_3}{\varphi_2}\right)},
 \eea
where the fields $\xi_{1,2,3}$ and $\theta_{1,2,3}$ represent the excitations around the vevs. As in the previous case, we can use a gauge transformation to remove a phase from 
the vev structure that ensures the absence of terms of the form appearing in eq.\ref{LkinExpansion}. The form of the vev in this case becomes,
\bea
\vev{ \Phi_1}&=&\varphi_1 e^{i\left(\si+\frac{\varphi_2}{\varphi_1}\ga\right)}, \nonumber\\
 \vev{\Phi_2}&=&\frac{1}{\sqrt{2}}(\varphi_1^2+\varphi_2^{2})^{1/2}e^{i\si},\\
 \vev{\Phi_3}&=&\varphi_2 e^{i\left(\si-\frac{\varphi_1}{\varphi_2}\ga\right)},\nonumber
\eea
where $\si$ and $\ga$ represent two independent phases\footnote{Again we have applied the limit $\dot{\varphi}\ll{\varphi\dot{\si}, \varphi\dot{\ga}}$. If we do not apply 
this limit the gauge transformation parameter ($\la$) needed to remove the linear term in $A_0$ can be found by integrating the coefficient of the $A_0$ term with respect to time. 
This is in general complicated and we choose to assume that $\varphi$ is varying very slowly with time.}.

In the unitary gauge, a form that preserves the canonically normalized kinetic terms reads,
\bea
\nonumber
\label{fields2fdexcitationsNoNG}
\Phi_1&=&(\varphi_1+\xi_1)e^{i\left(\si+\frac{\varphi_2}{\varphi_1}\ga+\frac{\theta\varphi_2+
\theta^\prime\varphi_1(2/3)^{1/2}}{\varphi_1(\varphi_1^2+\varphi_2^{2})^{1/2}}\right)}, \nonumber \\
\Phi_2&=&\left(\frac{1}{\sqrt{2}}(\varphi_1^2+\varphi_2^{2})^{1/2}+\xi_2\right)e^{i\left(\si+
\frac{\theta^\prime(2/3)^{1/2}}{(\varphi_1^2+\varphi_2^{2})^{1/2}}\right)}, \\ 
\Phi_3&=&(\varphi_2+\xi_3)e^{i\left(\si-\frac{\varphi_1}{\varphi_2}\ga-
\frac{\theta\varphi_1-\theta^\prime\varphi_2(2/3)^{1/2}}{\varphi_2(\varphi_1^2+\varphi_2^{2})^{1/2}}\right)},\nonumber
\eea
where $\xi_{1,2,3}$, $\theta$ and $\theta^\prime$ label the physical excitations around the vev once the NG boson has been gauged away. The resulting spectrum consists of one 
Higgs field with mass $M_1^2=3g^2(\varphi_1^2+\varphi_2^2)$, and four massless scalar fields.

We proceed, as before, by diagonalizing the kinetic terms and evaluating the $J$ matrix given in eq.\ref{jmatrix}. The non-vanishing entries of the $J$-matrix are
\begin{equation}
J_{1, 2}=J_{2, 1}=\frac{k-\sqrt{k^2+M_1^2}}{2\sqrt{3k}(k^2+M_1^2)^{1/4}}\dot{\ga},
\end{equation}
which demonstrates that in this case preheating can take place provided that $\dot{\ga}\ne 0$.

It is instructive to compare the two cases considered thus far. The first flat direction contained a single vev amplitude, the second contained two independent vev amplitudes. The 
final result, however, is the same for both cases. This demonstrates a simple property of flat direction vev decay: the determining factor is not the \emph{number of flat directions} 
present in the system, but the \emph{number of fields} that have vevs. In particular, a necessary (but not sufficient) condition for non-perturbative production of particles is the 
existence of at least one relative physical and dynamical phase between the field vevs that constitute the flat direction.

\section{Dynamics of the vev phases}

We now demonstrate that both physical phases are in general dynamical. In our particular toy example the cancellation of $U(1)^3$ and mixed $U(1)$-gravitational anomalies 
requires that we extend the field content of our model by including three additional complex superfields, $\Phi_4$, $\Phi_5$ and $\Phi_6$. We assign the $U(1)$ charges and 
$R$-Parity ($R_p$) as follows: 
\begin{center}
\begin{tabular}{|c|cccccc|} 
\hline 
 &$\Phi_1$&$\Phi_2$&$\Phi_3$&$\Phi_4$&$\Phi_5$&$\Phi_6$ \\ 
\hline
$U(1)$ & 1/2 & -1 & 1/2 & -1/2 &1 &-1/2\\
$R_p $&+&-&+&-&+&-\\
\hline
\end{tabular} 
\end{center}
This choice of $R_p$ assignments forbids the superpotential term $\Phi_1\Phi_2\Phi_3$, thus preserving F-flatness. There exist several possible flat directions for this particular 
field content. We assume that vevs for only $\Phi_1, \Phi_2$ and $\Phi_3$ are turned on, leaving $\Phi_4$, $\Phi_5$ and $\Phi_6$ with no vevs. With this assumption, the 
lowest dimension gauge invariant operators which have vevs are
\begin{equation}
\mathcal{O}_1=\Phi_1\Phi_2\Phi_3,\quad \mathcal{O}_2=\Phi_1^2\Phi_2 \quad \textrm{and}\quad \mathcal{O}_3=\Phi_2\Phi_3^2.
\end{equation}
Note that the last two operators depend on both physical phases, $\si$ and $\ga$.  
Using these operators, the lowest dimension terms appearing in the scalar potential, which are $R_p$ invariant and phase-dependent, arise as soft SUSY breaking A-terms and 
appear as
\begin{equation}
V\supset \sum_{i,j}\frac{A_{ij} m_{s}}{M^3}\mathcal{O}_i \mathcal{O}_j+h.c.,
\end{equation}
where $M$ denotes the cut off scale of the theory (e.g. the Planck mass or GUT scale), $m_{s}$ represents the scale of the SUSY breaking, and  $A_{ij}$ label 
dimensionless coefficients of order one. Lower order phase-independent interactions will also contribute to the lifting of the flat direction and have the generic forms
\begin{equation}
V\supset \sum_{i}\frac{m_i^2}{2}|\Phi_i|^2+\sum_{i,j}\frac{\lambda_{ij}}{8}|\Phi_i|^2|\Phi_j|^2,
\end{equation}
where $m_i^2$ denote the soft SUSY breaking masses, and the second terms arise from loop corrections with $\lambda_{i,j}\sim g^4 m_{s}^2/\varphi^2$ (see for example 
\cite{Affleck:1984fy} for similar loop induced terms). The potential for the single flat direction amplitude case considered in section \ref{sfd}, using the vev form shown in 
eq.\ref{mod1vev}, becomes
\begin{align}
V\supset\frac{m_1^2+m_2^2+m_3^2}{2}\varphi^2+\frac{\lambda^\prime}{4}\varphi^4
+\varphi^6\big(A^\prime_{11}e^{i6\si}+\nonumber\\ A^\prime_{12}e^{i(6\si+2\gamma)}+A^\prime_{13}e^{i(6\si-2\gamma)}
+A^\prime_{22}e^{i(6\si+4\gamma)}\\+A^\prime_{33}e^{i(6\si-4\gamma)}\big)\nonumber,
\end{align}
where $A^\prime_{ij}$ and $\lambda^\prime$ denote combinations of the couplings discussed above. The potential for the multiple vev amplitude case will be very similar with 
the obvious changes of vev amplitudes. The phase-dependent terms in the potential provide non-trivial dynamics for the phases $\si$ and $\ga$ and will in general lead to 
$\dot{\gamma}\neq 0$ and therefore a non-vanishing $J$ matrix. As discussed above, the appearance of a non-vanishing $J$ matrix can lead to the non-perturbative production of 
particles by the rotating flat direction: the condensate can decay via preheating.

\section{Two independent flat directions}\label{tfd}

A further instructive toy model consists of two independent flat directions existing simultaneously. We consider four chiral superfields $\Phi_1, \Phi_2, \Phi_3$, and $\Phi_4$ 
charged under a gauged $U(1)$ symmetry with charges $\pm q_1$ and $\pm q_2$ respectively. The potential arising from the D-terms reads
\beq
V=\frac{g^2}{8}(q_1\abs{\Phi_1}^2-q_1\abs{\Phi_2}^2+q_2\abs{\Phi_3}^2-q_2\abs{\Phi_4}^2)^2.
\label{potu1}
\eeq
Although this toy model has been examined previously in \cite{Olive:2006uw}, applying the methods outlined in the first sections of this letter helps establish the important 
properties of the model. The potential in eq.\ref{potu1} admits flat direction vevs of the following forms
 \bea
  \vev{\Phi_1}&=&\varphi_1e^{i\tilde{\si}_1},\nonumber \\
 \vev{\Phi_2}&=&\varphi_1e^{i\tilde{\si}_2},\\
 \vev{\Phi_3}&=&\varphi_2e^{i\tilde{\si}_3}, \nonumber\\
 \vev{\Phi_4}&=&\varphi_2e^{i\tilde{\si}_4}.\nonumber
 \eea
 We can write the excitations around the vevs as 
 \bea
 \nonumber
 \label{fields2}
 \Phi_1&=&(\varphi_1+\xi_1)e^{i(\tilde{\si}_1+\frac{\theta_1}{\varphi_1})},\nonumber \\
 \Phi_2&=&(\varphi_1+\xi_2)e^{i(\tilde{\si}_2+\frac{\theta_2}{\varphi_1})},\\
 \Phi_3&=&(\varphi_2+\xi_3)e^{i(\tilde{\si}_3+\frac{\theta_3}{\varphi_2})}, \nonumber\\
 \Phi_4&=&(\varphi_2+\xi_4)e^{i(\tilde{\si}_4+\frac{\theta_4}{\varphi_2})}.\nonumber
 \eea
As before we can make a gauge transformation and remove one phase in such a way that terms of the form shown in eq.\ref{LkinExpansion} vanish. The final form appears as
 \bea\label{4fieldvev}
  \vev{\Phi_1}&=&\varphi_1e^{i(\si_1+\gamma\frac{\varphi_2}{q_1\varphi_1})},\nonumber \\
 \vev{\Phi_2}&=&\varphi_1e^{i(\si_1-\gamma\frac{\varphi_2}{q_1\varphi_1})},\\
 \vev{\Phi_3}&=&\varphi_2e^{i(\si_2+\gamma\frac{\varphi_1}{q_2\varphi_2})}, \nonumber\\
 \vev{\Phi_4}&=&\varphi_2e^{i(\si_2-\gamma\frac{\varphi_1}{q_2\varphi_2})},\nonumber
 \eea
demonstrating the existence of three physical phases. Transforming in to the unitary gauge, we can write the excitations around the vevs as
 \bea
 \nonumber
 \label{fields2NoNG}
 \Phi_1&=&(\varphi_1+\xi_1)e^{i(\si_1+\gamma\frac{\varphi_2}{q_1\varphi_1}+
 \theta\frac{1}{\sqrt{2}\varphi_1}+\theta^{\prime\prime}\frac{q_2\varphi_2}{\varphi^{\prime}\varphi_1})},\nonumber \\
 \Phi_2&=&(\varphi_1+\xi_2)e^{i(\si_1-\gamma\frac{\varphi_2}{q_1\varphi_1}+\theta\frac{1}{\sqrt{2}\varphi_1}-
 \theta^{\prime\prime}\frac{q_2\varphi_2}{\varphi^{\prime}\varphi_1})},\\
 \Phi_3&=&(\varphi_2+\xi_3)e^{i(\si_2-\gamma\frac{\varphi_1}{q_2\varphi_2}+\theta^\prime\frac{1}{\sqrt{2}\varphi_2}-
 \theta^{\prime\prime}\frac{q_1\varphi_1}{\varphi^{\prime}\varphi_2})}, \nonumber\\
 \Phi_4&=&(\varphi_2+\xi_4)e^{i(\si_2+\gamma\frac{\varphi_1}{q_2\varphi_2}+\theta^\prime\frac{1}{\sqrt{2}\varphi_2}+
 \theta^{\prime\prime}\frac{q_1\varphi_1}{\varphi^{\prime}\varphi_2})},\nonumber
 \eea
where $\varphi^{\prime}=\sqrt{2}(q_1\varphi_1^2+q_2\varphi_2^2)^{1/2}$. The spectrum in this case consists of one massive Higgs particle and six massless scalar fields (the 
NG has been gauged away). Again, we must diagonalize the kinetic terms. Applying the necessary field redefinitions we are able to evaluate the $J$ matrix. The non-vanishing $J$ 
matrix elements read
 \bea
&& J_{1, 2}=J_{2, 1}=\frac{k-\sqrt{k^2+M_1^2}}{\sqrt{k}(k^2+M_1^2)^{1/4}}\frac{q_1q_2}{\varphi^{\prime 2}}(\dot{\si}_1-\dot{\si}_2)
\varphi_1\varphi_2\nonumber\\
&& J_{1, 3}=J_{3, 1}=\frac{k-\sqrt{k^2+M_1^2}}{\sqrt{2k}(k^2+M_1^2)^{1/4}}\frac{\dot{\gamma}\varphi_2}{\varphi^{\prime}}\\
&& J_{1, 4}=J_{4, 1}=\frac{k-\sqrt{k^2+M_1^2}}{\sqrt{2k}(k^2+M_1^2)^{1/4}}\frac{\dot{\gamma}\varphi_1}{\varphi^{\prime}},\nonumber
\eea
which depend on the relative phases between the field vevs. We should point out that only the Higgs eigenstate ($i=1$) is distinguishable. The other indices label the light fields 
which at this level are all massless. Preheating is again possible provided two of the phases have non-zero time derivatives. Using the particular case with $q_1=q_2$, we can write 
the scalar potential (see Appendix for details) yielding the terms
\begin{multline}\label{V4}
V=\frac{1}{2}(m_1^2+m_2^2)\varphi_1^2+\frac{1}{2}(m_3^2+m_4^2)\varphi_2^2+\frac{A_1}{8}\frac{m_{s}}{M}\varphi_1^4e^{i4\si_1}\\+\frac{A_2}{8}\frac{m_{s}}{M}\varphi_2
^4e^{i4\si_2}
+\frac{A_3}{8}\frac{m_{s}}{M}\varphi_1^2\varphi_2^2e^{i2(\si_1+\si_2+\ga\frac{\varphi_2^2-\varphi_1^2}{\varphi_2\varphi_1})}\\+\frac{A_4}{8}\frac{m_{s}}{M}\varphi_1^2
\varphi_2^2e^{2i(\si_1+\si_2-\ga\frac{\varphi_2^2-\varphi_1^2}{\varphi_2\varphi_1})}+... 
\end{multline}
Clearly, non-trivial dynamics exist for the phases $\gamma$, $\si_1$ and $\si_2$.

\section{Numerical analysis}

As a proof-of-principle that achieves quantitative results, we numerically analyse the model described in section \ref{tfd}. We use a simplified version of the potential appearing in 
eq.\ref{V4}, confining ourselves to the potential
\begin{multline}
V = \frac{1}{2} m_{\varphi_1}^2 \varphi_1^2 + \frac{1}{2} m_{\varphi_2}^2 \varphi_2^2 + 
\frac{A_1}{8}\frac{m_{s}}{M}\varphi_1^4e^{i4\si_1}\\+\frac{A_2}{8}\frac{m_{s}}{M}\varphi_2^4e^{i4\si_2}+\mbox{h.c}
\end{multline}
where $m_{\varphi_1}^2=m_1^2+m_2^2$,  $m_{\varphi_2}^2=m_3^2+m_4^2$. This potential decouples the equations of motion for $\ga$, $\si_1$ and $\si_2$. The equation of 
motion for $\ga$ reduce to $\ddot{\ga}=0$, and with the choice of initial conditions, $\dot{\ga}=0$, the effects of  $\ga$ on preheating are removed. Our simplified potential 
allows us to numerically evolve the classical evolution of the flat direction vevs and analyze particle production in a self-consistent background. We also make the simplifying 
assumption setting $A_1 = A_2 =\la \frac{M}{m_s}$ in eq.\ref{V4}. Again, we stress that we use this grossly simplified potential simply to demonstrate the quantitative behaviour 
of the toy model class.

Measuring the conformal time in units of $\tau \rightarrow f t$ with $ f= g\varphi_{1_0}$ and using the re-scaled flat-direction vev amplitudes
\bea
\varphi_1 &=& \frac{\varphi_{1_0}}{a} F_1\nonumber \\
\varphi_2 &=& \frac{\varphi_{2_0}}{a} F_2
\eea
we find the background equations,
\begin{multline}\label{F}
F_1^{\prime\prime} +\left[\frac{\mu_1^2 a^2}{2} - \sigma_1^{\prime 2} - \frac{a^{\prime\prime}}{a}\right] F_1 + \frac{\la F_1^3 }{2 g^2} \cos\left(4\sigma_1\right) = 0  \\
F_2^{\prime\prime} +\left[\frac{\mu_2^2 a^2}{2} - \sigma_2^{\prime 2} - \frac{a^{\prime\prime}}{a}\right] F_2 + \frac{\la F_2^3 
}{2g^2}\left(\frac{\varphi_{2_0}}{\varphi_{1_0}}\right)^2\cos\left(4\sigma_2\right) =0
\end{multline}
where a prime represents a derivative with respect to $\tau$ and
\bea\label{sigmadot}
&& \sigma_1^{\prime\prime} + 2 \sigma_1^\prime \frac{F_1^\prime}{F_1} - \frac{\lambda}{2 g^2} F_1^2 \sin(4\sigma_1) = 0 \nonumber \\
&& \sigma_2^{\prime\prime} + 2 \sigma_2^\prime \frac{F_2^\prime}{F_2} - \frac{\lambda}{2 g^2}\left(\frac{\varphi_{2_0}}{\varphi_{1_0}}\right)^2F_2^2 \sin(4\sigma_2) = 0
\eea
describes the motion of the flat direction vevs; $\mu_1 = m_{\varphi_1}/f, \mu_2 = m_{\varphi_2}/f$. The scale factor evolves as,
\begin{multline}
\frac{a^{\prime\prime}}{a} =-\frac{a^{\prime 2}}{a^2} + \frac{1}{2}\left[f_p^2\left\{\mu_1^2 F_1^2 + \frac{\lambda}{2g^2}\frac{F_1^4}{a^2}\cos(4\sigma_1)\right.\right. \\
+ \left.\left.\mu_2^2 \left(\frac{\varphi_{2_0}}{\varphi_{1_0}}\right)^2 F_2^2 + 
\frac{\lambda}{2g^2}\left(\frac{\varphi_{2_0}}{\varphi_{1_0}}\right)^4\frac{F_2^4}{a^2}\cos(4\sigma_2)\right\} + \frac{a^2\rho_\psi}{M^2_{pl} f^2}\right]
\end{multline}
where $\rho_\psi$ is the energy density of the inflaton field and $f_p = \varphi_{1_0}/M_{pl}$ is set to $f_p = 0.1$ in our numerics. We also take $\mu_1 =10^{-2}, \mu_2
=10^{-2}/2$, and $\lambda = \mu_1^2$ for computational ease. As initial conditions, we start the flat direction at rest, such that $(\varphi_{1,2}\exp(i\sigma_{1,2}))^\prime=0$. 
We choose to set initially $F_{1,2} =1$, $\sigma_{1,2} = 0.05$, $\sigma_{1,2}^\prime =0$, $a=1$ and  $a^\prime = \mu_1$, which implies $F_{1,2}^\prime = a^\prime = \mu_1$. 
While these initial conditions do not present a realistic case (where $\mu\sim 10^{-14}$ and $F_{1,2}^\prime \gg \mu_1$), they do provide a numerical proof-of-principle similar 
to \cite{Olive:2006uw}.
\begin{figure}[ht!]
\newlength{\picwidtha}
\setlength{\picwidtha}{3.7in}
 \begin{flushleft}
{\resizebox{\picwidtha}{!}{\includegraphics{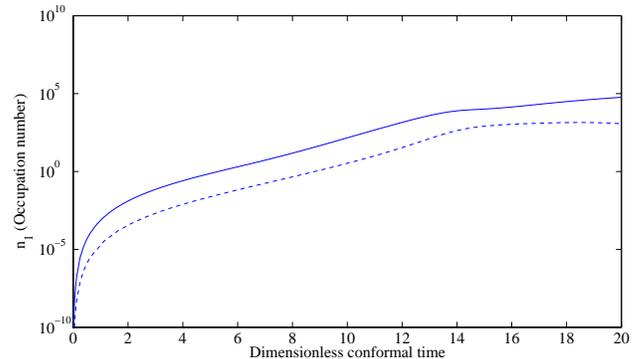}}}
 \end{flushleft}
\caption{Occupation number for one of the excited fields as a function of dimensionless conformal time, obtained using eq.\ref{n} after numerically integrating the background 
field equations and Bogolyubov matrices; $k=\mu_1/3\times10^{5}$, other parameters as explained in the text. The solid lines represents preheated fields with 
$\varphi_{2_0}/\varphi_{1_0} = 1$ while the dashed lines indicate the preheated fields with $\varphi_{2_0}/\varphi_{1_0} = 0.1$}
\label{ratios}
\end{figure}

Initially the flat direction vevs correspond to a condensate of coherent particles with vanishing momentum. The motion of these vevs, described by eq.\ref{F} and 
eq.\ref{sigmadot}, and the interactions described in the previous section, cause the rapid decay of this condensate into a decoherent state of particles. FIG. 1 shows the occupation 
numbers, $n_i(t)$, of these light particles as a function of conformal time: the exponential growth of these functions signals the exponentially fast decay of the flat-direction vev. 
The two line types, solid and dashed, represent the ratios $\varphi_{2_0}/\varphi_{1_0} =1, 0.1$ respectively. We see that preheating occurs over a wide range of the ratio 
$\varphi_{2_0}/\varphi_{1_0}$. In this numerical example, preheating effects do not vanish until $\varphi_{2_0}/\varphi_{1_0} \lesssim 10^{-2}$.
FIG 2 displays the resulting spectrum for one of the light fields, we see that production of higher momentum modes becomes kinematically suppressed.
\begin{figure}[ht!]
\newlength{\picwidthc}
\setlength{\picwidthc}{3.7in}
 \begin{flushleft}
{\resizebox{\picwidthc}{!}{\includegraphics{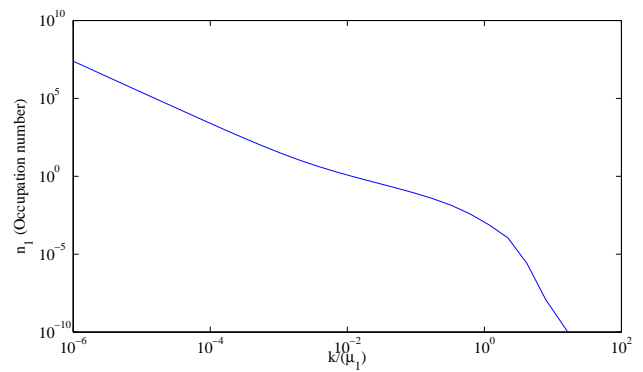}}}
 \end{flushleft}
\caption{Occupation number as a function of comoving momentum obtained as for FIG. 1, with $\varphi_{2_0}/\varphi_{1_0} = 0.5$ at time $t=20$. }
\label{ratios1}
\end{figure}

We must stress that effects of SUSY breaking terms in the Lagrangian eq.\ref{Lagrangian} will significantly affect the amount of particle production produced by the rotating 
condensate. A mass term for the light fields translates into a momentum shift in eq.\ref{Omega} and this corresponds to a kinematic suppression of the modes 
\cite{Olive:2006uw}. 
A realistic model involving MSSM flat directions will in general contain many SUSY breaking terms and non-renormalisable operators, creating large model dependencies in the 
precise determination of the momentum shift.   Consequently, we leave such a study to future work.

\section{Conclusions}

The cosmological fate of flat directions provides a major ingredient for the history of the early Universe. Flat directions can provide mechanisms for generating the baryon 
asymmetry of the Universe and can play an important role in reheating after inflation. Our analysis stresses the use of the unitary gauge in which the physical content of the 
theory becomes manifest. By transforming to the unitary gauge, complications arising from massless NG modes in the mixing of the excitations around the flat direction vev are 
removed. The mixing matrix in this gauge defines the mass eigenstates of the physical scalar fields and determines if non-perturbative decay is possible. Since the mass matrix in 
the unitary gauge can contain time dependent mixing among all fields, one of the necessary conditions for preheating can be satisfied.

Two further crucial conditions for preheating in our analysis center on the existence of physical relative phases between the field vevs that make up the flat direction(s) and that 
these phases possess non-trivial dynamics during the early universe. The first of these conditions generally becomes satisfied if the difference between the number fields that 
acquire a vev and the number of broken diagonal generators is larger than one -- every diagonal generator removes one unphysical phase. The second condition generally becomes 
satisfied if terms which explicitly depend on the phase differences appear in the scalar potential. The existence of gauge invariant products of background fields 
exhibiting this phase dependence represents the crucial ingredient and determines the phase dependence in the scalar potential. Once these conditions are satisfied, the flat 
direction condensate can decay non-perturbatively via preheating.

\section{Appendix}

In this appendix we specify a simple model to justify the form of the potential given in eq.\ref{V4}. We have the following field content with $U(1)$ charges and $R_p$ 
assignments, \vspace{-3mm}
 \begin{center} \begin{tabular}{|c|cccc|} \hline &$\Phi_1$&$\Phi_2$&$\Phi_3$&$\Phi_4$\\
\hline
$U(1)$ & 1 & -1 & 1 & -1 \\
$R_p $&-&+&-&+\\
\hline
\end{tabular}
\end{center}

The lowest dimension gauge invariant operators are \bea\label{operators4}
\mathcal{O}_1&=\Phi_1\Phi_2,\quad \mathcal{O}_2&=\Phi_3\Phi_4,\nonumber\\
 \mathcal{O}_3&=\Phi_1\Phi_4,\quad\mathcal{O}_4&=\Phi_2\Phi_3.
\eea
The lowest dimension terms which are $R_p$ invariant and phase-dependent arise as soft SUSY breaking A-terms and appear in the scalar potential as 
\begin{multline} 
V=\sum_{i=1}^4 \frac{m_i^2}{2}|\Phi_i|^2+\frac{A_1}{8}\frac{m_{s}}{M}\mathcal{O}_1^2+\frac{A_2}{8}\frac{m_{s}}{M}\mathcal{O}_2^2\\
+\frac{A_3}{8}\frac{m_{s}}{M}\mathcal{O}_3^2+\frac{A_4}{8}\frac{m_{s}}{M}\mathcal{O}_4^{2}+...
\end{multline}
where the ellipses stand for other terms of the same order with different products of the gauge invariant operators, loop induced contributions and higher order terms in the 
$(1/M)$ expansion. Substituting the vevs given in eq.\ref{4fieldvev} we generate the potential terms given in eq.\ref{V4}.

\vspace{-5mm}

\section{Acknowledgements}

We thank R. Allahverdi and J. March-Russell for many useful discussions. We are also grateful to I. Aitchison, S. Hannestad, P. Iafelice, G. Ross, S. Sarkar, D. Ghilencea, M. 
Schvellinger and M. Sloth for helpful discussions. AB wishes to thank Oxford University for hospitality during the course of this work, DM is supported by the NSERC Canada, FR 
by the Greendale scholarship. This work was partially supported by the EU 6th Framework Marie Curie Research and Training network ``UniverseNet" (MRTN-CT-2006-035863).

\end{document}